\documentclass[prb,twocolumn,showpacs,groupedaddress]{revtex4}
\usepackage{amssymb,mathptm,dcolumn,bm,times,graphicx,color}

\begin{document}

\title{NMR relaxation rate in superconducting pnictides: extended $s_{\pm}$ scenario}

\author{D.~Parker$^{1,2,6}$}
\author{O.V.~Dolgov$^3$}
\author{M.M.~Korshunov$^{1,4}$}
\author{A.A.~Golubov$^{5}$}
\author{I.I.~Mazin$^{6}$}

\affiliation{$^1$Max-Planck-Institut f\"{u}r Physik komplexer Systeme, D-01187 Dresden,
Germany}
\affiliation{$^2$Max-Planck Institut f\"{u}r Chemische Physik fester Stoffe, D-01187
Dresden, Germany}
\affiliation{$^3$Max-Planck-Institut f\"{u}r Festk\"{o}rperforschung, D-70569 Stuttgart,
Germany}
\affiliation{$^4$L.V. Kirensky Institute of Physics, Siberian Branch of RAS, 660036
Krasnoyarsk, Russia}
\affiliation{$^5$Faculty of Science and Technology, University of Twente, 7500 AE
Enschede, The Netherlands}
\affiliation{$^6$Naval Research Laboratory, 4555 Overlook Ave. SW, Washington, DC 20375}
\date{\today}

\begin{abstract}
Recently, several measurements of the nuclear spin lattice relaxation rate
$T_{1}^{-1}$ in the newly discovered superconducting Fe-pnictides have been
reported. These measurements generally show no coherence peak below $T_{c}$ and
indicate a low temperature power law behavior, the characteristics commonly
taken as evidence of unconventional superconductivity with lines of nodes
crossing the Fermi surface. In this work we show that (i) the lack of a
coherence peak is fully consistent with the previously proposed nodeless
extended $s_{\pm}$-wave symmetry of the order parameter (whether in the clean
or dirty limit) and (ii) the low temperature power law behavior can be also
explained in the framework of the same model, but requires going beyond the
Born model.
\end{abstract}

\pacs{74.20.Rp, 76.60.-k, 74.25.Nf, 71.55.-i}
\maketitle

The recently synthesized \cite{kamihara}, high-$T_{c}$ superconducting
ferropnictides may be the most enigmatic superconductors discovered so far. One
of the biggest mysteries associated with these materials is that now, with
improved sample quality and single crystal availability, some experiments
unambiguously see a fully gapped superconducting state
\cite{PCAR1,PCAR2,PCAR3,ARPES1,ARPES2,ARPES3,pd1,pd2,pd3} and an $s$-wave
pairing \cite{SQUID}, while others unequivocally point towards line nodes in
the gap \cite{NMR1,NMR2,NMR3,NMR4}. Particularly disturbing is that on both
sides data of high quality were reported, by highly reputable groups, so
experimental errors seem unlikely. It is possible that full reconciliation will
require a highly advanced theory that will treat both superconducting and spin
density wave order parameters on equal footing, and include the interaction
between the two. Nevertheless, it is interesting, and important, to investigate
more conventional options first.

Evidence for fully gapped superconductivity comes from three different sources:
Andreev reflection \cite{PCAR1,PCAR2,PCAR3}, exponential temperature dependence
of the penetration depths \cite{pd1,pd2,pd3}, and the angle-resolved
photoemission spectroscopy (ARPES) \cite{ARPES1,ARPES2,ARPES3}. That three so
different probes yield qualitatively the same result is very convincing. Yet
the nuclear magnetic resonance (NMR) spin-lattice relaxation rate, $1/T_{1}$,
does not show two classical fingerprints of a conventional fully-gapped
superconductors: the Hebel-Slichter coherence peak and the exponential decay at
low temperature, but, rather, a power-like law \cite{NMR1,NMR2,NMR3,NMR4},
usually referred to as $T^{3}$, but in reality somewhere between $T^{3}$ and
$T^{2.5}$. Such behavior is usually taken to be an evidence for a $d$-wave or
similar superconducting state with lines of nodes. However, it was pointed out
\cite{Ishida} that in dirty $d$-wave samples at low temperatures behavior
changes from $T^{3}$ to $T$ (as node lines are washed out into node spots by
impurities), and this was not observed in ferropnictides.

So far the evidence in favor of nodeless superconductivity seems stronger.
Therefore, it is interesting to check whether it may be possible to explain
the results of the NMR experiments without involving an order parameter with
node lines.
\begin{figure}[tbp]
\includegraphics[width=0.5\linewidth]{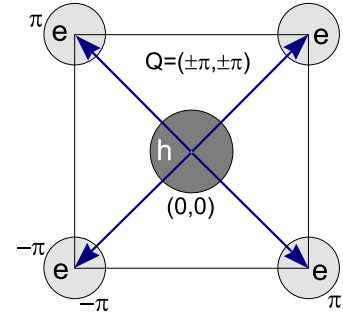}
\caption{(color online) A depiction of the Fermi surface geometry for the
Fe-based oxypnictides. Hole and electron Fermi surface pockets are
indicated, and an antiferromagnetic wave vector $\mathbf{Q}$ is also shown.}
\label{fig1}
\end{figure}

In this paper we calculate $1/T_{1}T$ for a model superconductor consisting of
two relatively small semimetallic Fermi surfaces, separated by a finite wave
vector $\mathbf{Q}$ (Fig.~\ref{fig1}). This is an approximation to the Fermi
surface of ferropnictides. We intentionally drop quantitative details that may
differ from compound to compound, and consider the simplest possible case with
the same densities of states on each surface \cite{gaps}. We further assume
that each surface features the same gap \cite{gaps}, but the relative phase
between the two order parameters is $\pi$. This is the so-called $s_{\pm}$
model, proposed in Ref.~\onlinecite{mazin} and discussed in
Refs.~\onlinecite{kuroki,korna1,korna2} and a number of more recent
publications by various groups. In the spirit of this model and of model
calculations \cite{kuroki,korna1,korna2} we will assume that the total
(renormalized) spin susceptibility is strongly peaked at and around
$\mathbf{Q}$.

We will show here that in this model the Hebel-Slichter peak is strongly
suppressed already in the clean limit and can be entirely eliminated even by a
very weak impurity scattering. On the contrary, the low temperature behavior
remains exponential even in the strong coupling limit. Introducing impurities
does create strong deviations from the exponential behavior, but in the Born
approximation the effect is stronger just below superconducting temperature
$T_{c}$ and weaker at $T \rightarrow 0$, so that the observed power law
behavior (down to at least $0.1 T_{c}$) is very difficult to reproduce. This
behavior, however, can be reproduced if one goes beyond the Born limit of
impurity scattering.

The NMR relaxation rate, assuming a Fermi contact hyperfine interaction
\cite{hf}, is given by the standard formula: $({1}/{T_{1}T) \propto
\lim_{\omega \rightarrow 0} \sum_{\mathbf{q}}
\mathrm{Im}\chi_{\pm}(\mathbf{q},\omega)}/{\omega},$ where
$\chi_{\pm}(\mathbf{q},\omega)$ is the analytic continuation of the Fourier
transform of the correlation function
$\chi_{\pm}(\mathbf{r},\tau)=-\left\langle \left\langle
T_{\tau}S_{+}(\mathbf{r},-\mathrm{i}\tau)S_{-}(\mathbf{0},0)\right\rangle
\right\rangle_{imp}$, averaged (if needed) over the impurity ensemble. Here,
$S_{\pm}(\mathbf{r},-\mathrm{i}\tau)=\exp(H\tau)S_{\pm}(\mathbf{r})\exp(-H\tau)$,
where $H$ is the electronic Hamiltonian, $\tau $ denotes imaginary time, and
$S_{\pm}$ is expressed via the electron operators as
$S_{+}(\mathbf{r})=\psi_{\uparrow}^{\dag}(\mathbf{r})\psi_{\downarrow}(\mathbf{r})$
and
$S_{-}(\mathbf{r})=\psi_{\downarrow}^{\dag}(\mathbf{r})\psi_{\uparrow}(\mathbf{r})$.
Adopting the above-described model, we can keep only the interband contribution
to $\chi$, in which case this formula simplifies to
\begin{equation}
1/T_{1}T \propto \lim_{\omega \rightarrow 0}\mathrm{Im}\chi_{12}(\omega)/\omega,
\end{equation}
where $\chi_{12}(\omega)$ is obtained by integrating over all $\mathbf{q}$'s
connecting the two Fermi surfaces (obviously, only $\mathbf{q} \sim \mathbf{Q}$
contribute). In case of a weakly coupled clean superconductor below $T_{c}$ we
have
\begin{equation}
\frac{1}{T_{1}T} \propto \sum_{\mathbf{kk^{\prime}}} \left(1+\frac{\Delta_{1}\Delta_{2}}
{E_{\mathbf{k}}E_{\mathbf{k^{\prime}}}} \right)
\left[ -\frac{\partial f(E_{\mathbf{k}})}{\partial E_{\mathbf{k}}} \right]
\delta \left( E_{\mathbf{k}}-E_{\mathbf{k^{\prime}}} \right),
\end{equation}
where $\mathbf{k}$ and $\mathbf{k^{\prime}}$ lie on the hole and the electron
Fermi surfaces, respectively, $E_{\mathbf{k}}$ is the quasiparticle energy in
the superconducting state, $\Delta_{1}$ and $\Delta_{2}$ are the
superconducting gaps on hole and electron Fermi surfaces, and $f(E)$ is the
Fermi distribution function. This is a straightforward generalization of the
textbook expression \cite{schrieffer}. Following the usual BCS prescription,
$\sum_{k} \rightarrow \int_{\Delta}^{\infty}E dE/\sqrt{E^{2}-\Delta^{2}}$, the
$\mathbf{k}$-space sum can be converted to an energy integral and for a
conventional $s$-wave superconductor with $\Delta_{1}=\Delta_{2}=\Delta$ one
finds
\begin{equation}
\frac{1}{T_{1}}\propto \int\limits_{\Delta (T)}^{\infty }dE
\frac{E^{2}+\Delta^{2}}{E^{2}-\Delta^{2}}\mathrm{sech}^{2} \left(\frac{E}{2T}\right).
\end{equation}
The denominator gives rise to a peak just below $T_{c}$, the famous
Hebel-Slichter peak. As pointed out in Ref.~\onlinecite{mazin}, it is
suppressed for the $s_{\pm}$ state. Indeed, if $\Delta_{1}=-\Delta_{2}=\Delta$,
\[
\frac{1}{T_{1}} \propto \int\limits_{\Delta(T)}^{\infty}dE
\frac{E^{2}-\Delta^{2}}{E^{2}-\Delta^{2}} \mathrm{sech}^{2} \left( \frac{E}{2T} \right)
= \int\limits_{\Delta(T)}^{\infty}dE \mathrm{sech}^{2} \left( \frac{E}{2T} \right).
\]
As $T$ decreases from $T_{c}$, the integral decreases monotonically.

In a more general case, when $\Delta_{1}\Delta_{2} < 0$ and $|\Delta_{1}| \neq
|\Delta_{2}|$,
\begin{equation}
\frac{1}{T_{1}T} \propto
\int\limits_{\max \left\{ |\Delta_{1}|,|\Delta_{2}|\right\} }^{\infty }d\varepsilon
\left( -\frac{\partial f(\varepsilon)}{\partial \varepsilon}\right)
\frac{\varepsilon^{2}-|\Delta_{1}\Delta_{2}|}
{\sqrt{\varepsilon^{2}-\Delta_{1}^{2}}\sqrt{\varepsilon^{2}-\Delta_{2}^{2}}}.
\end{equation}
Following Fibich~\cite{fibich1,fibich2}, we assume $\left( -{\partial
f(\varepsilon)}/{\partial \varepsilon} \right)$ to be a slow varying function
and obtain (for $\Delta_{1} > |\Delta_{2}|$)
\begin{eqnarray*}
1/T_{1}T &\propto& f(\Delta_{1})+2I(\Delta_{1},\Delta_{2})f(\Delta_{1})
\left[ 1-f(\Delta_{1}) \right]/T, \\
I(\Delta_{1},\Delta_{2}) &=& \mathbf{K} \left( \Delta_{1} / \Delta_{2}\right)
\left( \Delta_{1}+\Delta_{2} \right) - \mathbf{E}\left( \Delta_{1} / \Delta _{2} \right)
\Delta_{1},
\end{eqnarray*}
where $\mathbf{K}(x)$ and $\mathbf{E}(x)$ are the complete elliptic integrals
of the first and second kind, respectively. When $\Delta_{1}=\Delta_{2}$, $I$
is reduced to the standard BCS formula, and when $\Delta_{1}=-\Delta_{2}$ it
vanishes identically.

Let us now include impurity scattering and move to the strong coupling limit.
Following Samokhin and Mitrovi\'{c}~\cite{SM,MS}, we can write down the
following formula:
\begin{eqnarray*}
\frac{1}{T_{1}T} &\propto& \int\limits_{0}^{\infty}d\omega
\left( -\frac{\partial f(\omega)}{\partial \omega} \right) \nonumber \\
&\times& \left\{ \left[ \mathrm{Re}g_{1}^{Z}(\omega)+\mathrm{Re}g_{2}^{Z}(\omega) \right]^{2}
+ \left[ \mathrm{Re}g_{1}^{\Delta}(\omega )+\mathrm{Re}g_{2}^{\Delta }(\omega) \right]^{2}
\right\}.
\end{eqnarray*}
For our model $g_{i}^{Z}(\omega)=n_{i} \left( \omega \right)
Z_{i}(\omega)\omega /D_{i}(\omega)$, $g_{i}^{\Delta}(\omega)=n_{i}\left( \omega
\right) \phi_{i}(\omega)/D_{i}(\omega)$, where $D_{i}(\omega )=\sqrt{\left[
Z_{i}(\omega)\omega \right]^{2}-\phi_{i}^{2}(\omega)}$, $Z_{i}(\omega)$ is the
mass renormalization, $\phi_{i}(\omega)=Z_{i}(\omega) \Delta_{i}(\omega)$ and
$n_{i} \left( \omega \right)$ is a partial density of states.

The renormalization function $Z_{i}(\omega)$ and complex order parameter
$\phi_{i}(\omega)$ have to be obtained by a numerical solution of the
Eliashberg equations. On the real frequency axis they have the form (we neglect
all instant contributions and consider a uniform impurity scattering with the
impurity potential $v_{ij}=v$),
\begin{eqnarray*}
\phi_{i}(\omega) &=& \sum_{j}\int\limits_{-\infty}^{\infty}dz
K_{ij}^{\Delta}(z,\omega)\mathrm{Re}g_{j}^{\Delta}(z)
+\mathrm{i}\gamma\frac{g_{1}^{\Delta}(\omega)-g_{2}^{\Delta}(\omega)}{2\mathcal{D}} \\
(Z_{i}(\omega)-1)\omega &=& \sum_{j}\int\limits_{-\infty}^{\infty}dz
K_{ij}^{Z}(z,\omega)\mathrm{Re}g_{j}^{Z}(z)
+\mathrm{i}\gamma\frac{g_{1}^{Z}(\omega)+g_{2}^{Z}(\omega)}{2\mathcal{D}},
\end{eqnarray*}
where $\mathcal{D}=1-\sigma+\sigma \left\{ \left[
g_{1}^{Z}(\omega)+g_{2}^{Z}(\omega) \right]^{2}-\left[
g_{1}^{\Delta}(\omega)-g_{2}^{\Delta}(\omega) \right]^{2} \right\}$, $\gamma=2
c\sigma/\pi N(0)$ is the normal state scattering rate, $N(0)$ is the the
density of states at the Fermi level, $c$ is the impurity concentration, and
$\sigma =\frac{\left[\pi N(0)v\right]^2} {1+\left[\pi N(0)v\right]^2}$ is the
impurity strength ($\sigma \rightarrow 0$ corresponds to the Born limit, while
$\sigma=1$ to the unitary one). Kernels $K_{ij}(z,\omega)$ are
\[
K_{ij}^{\Delta,Z}(z,\omega)=\int\limits_{0}^{\infty }d\Omega
\frac{\tilde{B}_{ij}(\Omega)}{2}
\left[ \frac{\tanh \frac{z}{2T}+\coth \frac{\Omega}{2T}}
{z+\Omega-\omega-\mathrm{i}\delta}-\left\{ \Omega \rightarrow - \Omega \right\} \right],
\]
where $\tilde{B}_{ij}(\Omega)$ is equal to $B_{ij}(\Omega)$ in the equation for
$\Delta $, and to $\left| B_{ij}(\Omega) \right|$ in the equation for $Z$. Note
that all retarded interactions enter the equations for the renormalization
factor $Z$ with a positive sign.

It is well known that pair-breaking impurity scattering dramatically increases
the subgap density of states just below $T_{c}$, and even weak magnetic
scattering can eliminate the Hebel-Slichter peak in conventional
superconductors. In our model, the same effect is present due to the
nonmagnetic interband scattering (the magnetic scattering, on the contrary, is
not pair-breaking in the Born limit) \cite{GolMaz}. Since the Hebel-Slichter
peak is not present in this scenario even in a clean sample, the pair-breaking
effect is more subtle: it changes exponential behavior below $T_{c}$ to a more
power-law like one (the actual power and extent of the temperature range with a
power-law behavior depend on the scattering strength). Note that in the Born
approximation the exponential behavior is always restored at low enough
temperature, unless the impurity concentration is so strong that $T_{c}$ is
suppressed by at least a factor of two \cite{GolMaz}.

Another well-known pair-breaking effect is scattering by thermally excited
phonons (or other bosons). This is, of course, a strong coupling effect. For
instance, strong coupling can nearly entirely eliminate the Hebel-Slichter peak
in a conventional superconductor \cite{DolgovGolubov,AllenRainer}. However,
this effect is even more attached to a temperature range just below $T_{c}$
since at low temperature boson excitations are exponentially suppressed.

Currently, most experimental data for ferropnictides go down in temperature to
$\sim 0.2-0.3 T_{c}$, but some results are available at temperatures as low as
$0.1 T_{c}$. So far exponential behavior has not been observed, which casts
doubt that Born impurity scattering may be responsible for such behavior.

Unitary scattering, on the other hand, has rather different low-temperature
behavior. As discussed, for instance, by Preosti and Muzikar~\cite{PM}, unitary
scattering in the case of $s_{\pm}$ superconducting gap (our choice of
$|\Delta_{1}|=|\Delta_{2}|$ corresponds to their parameter $r$ set to zero) the
subgap density of states is controlled by the unitarity parameter $\sigma$,
while the suppression of $T_{c}$ is controlled by a \textit{different}
parameter, namely, by the net scattering rate $\gamma$. The unitary limit
corresponds to $\sigma \rightarrow 1$, but $\gamma $ may be rather small at low
concentrations. The physical meaning is that here the diluted unitary limit
corresponds to the so-called ``swiss cheese'' model: each impurity creates a
bound state that contributes to the subgap density of states, but hardly to the
$T_{c}$ suppression. Indeed, Preosti and Mizikar have shown (see Fig.~1 in
Ref.~\onlinecite{PM}) that in this limit nonzero density of states at the Fermi
level appears already at zero temperature at arbitrary low impurity
concentration. That is to say, the bound state has zero energy. This is a
qualitatively new effect compared to the Born limit: In a dilute unitary regime
($\gamma \ll \Delta$), the NMR relaxation at $T \approx 0$ is mainly due to the
bound states at $E=0$; upon heating $1/T_{1}T$ initially remains constant, or
may even slightly decrease because if depopulating of the bound state. When the
temperature increases further, at some crossover temperature the relaxation
becomes dominated by thermal excitations across the gap and $1/T_{1}T$ starts
growing exponentially. When the gap is suppressed by the temperature as to
become comparable with $\gamma$, yet another effect kicks in, broadening of the
coherence peak near $T_{c}$ (less important for our $s_{\pm}$ state). Thus,
unitary scattering makes the $1/T_{1}T$ temperature dependence rather complex,
although the strongly unitary regime with low impurity concentration is rather
far from a power law behavior (even though being strongly nonexponential, see
Fig.~\ref{fig3}).

These qualitative arguments suggest that neither purely Born nor purely unitary
limits are well suited for explaining the observed $1/T_{1}$ behavior: the
former leads to an exponential behavior at low temperatures, the latter to
Korringa behavior. On the other hand, an \textit{intermediate} regime seems to
be rather promising in this aspect. Indeed, the energy of the above-mentioned
bound state is related to $\sigma$ as $E_{b}=|\Delta|\sqrt{1-\sigma}$. Thus, by
shifting $\sigma$ toward an intermediate scattering $\sigma \sim 0.5$ (which is
probably more realistic than either limit anyway) and increasing $\gamma$ we
create broad and monotonously increasing density of states, much closer to what
would be expected from the NMR data. It is worth noting that a
\textit{distribution} of $\sigma$'s (presence of different impurities with
different scattering strength), which we do not consider here, will also lead
to broadening of the bound state and work as an effectively enhanced $\gamma$
(therefore it is reasonable to try relatively large $\gamma$'s keeping in mind
that part of this is simulating).

We will now illustrate the above discussion using specific numerical models.
First, we present numerical solutions of the Eliashberg equations using a
spin-fluctuation model for the spectral function of the intermediate boson:
$B_{ij}(\omega)=\lambda_{ij}{\pi \Omega_{sf}}/({\Omega_{sf}^{2}+\omega^{2}})$,
with the parameters $\Omega_{sf}=25$ meV, $\lambda_{11}=\lambda_{22}=0.5$, and
$\lambda_{12}=\lambda_{21}=-2$. This set gives a reasonable value for $T_{c}
\simeq 26.7$ K. A similar model was used in Ref.~\onlinecite{timusk} to
describe optical properties of ferropnictides. The actual details of the
function are in fact not important; our usage of this particular function does
not constitute an endorsement or preference compared to other possibilities,
but is just used here for concreteness.
\begin{figure}[tbp]
\includegraphics[width=0.9\linewidth]{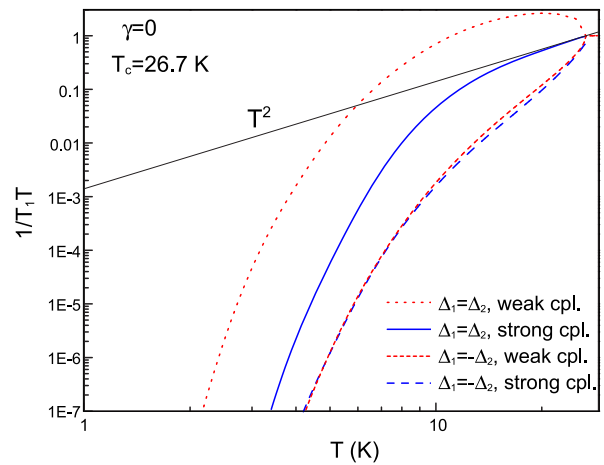}
\caption{(color online) Temperature dependence of the spin-lattice
relaxation rate $1/T_{1}T$ calculated in the clean limit ($\gamma=0$)
for a conventional $s$-wave superconductor ($\Delta_{1}=\Delta_{2}$) and
for an $s_{\pm}$ superconductor ($\Delta_{1}=-\Delta_{2}$). Results of
both weak and strong coupling approximations are shown.}
\label{fig2}
\end{figure}
\begin{figure}[tbp]
\includegraphics[width=0.99\linewidth]{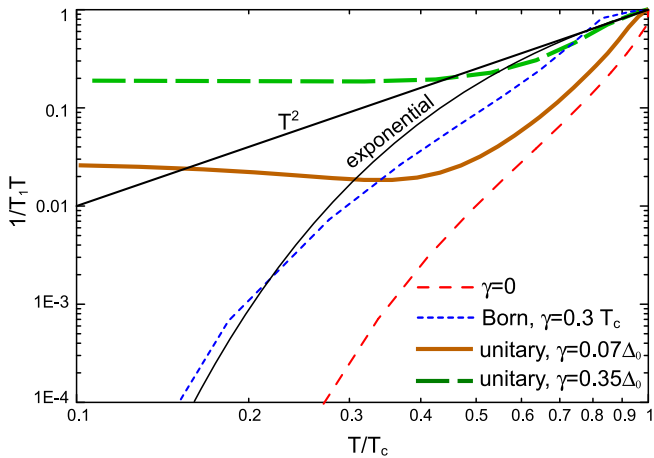}
\caption{(color online) Calculated temperature dependence of the spin-lattice
relaxation rate $1/T_{1}T$ for an $s_{\pm}$ superconductor in the strong coupling
approximation without impurities ($\gamma=0$),
with impurities in the Born limit ($\gamma=0.3 T_{c}$),
and in the unitary limit for small ($\gamma=0.07 \Delta_{0}$)
and large ($\gamma=0.35 \Delta_{0}$) impurity concentrations.}
\label{fig3}
\end{figure}

In Fig.~\ref{fig2} we compare the temperature dependence of the relaxation rate
calculated as described above in the clean limit for a conventional $s$-wave
superconductor ($\Delta_{1}=\Delta_{2}$) and for an $s_{\pm}$ superconductor
($\Delta_{1}=-\Delta_{2}$), both in the weak and in the strong coupling limits.
We observe that while in the conventional case strong coupling makes a big
difference by suppressing the coherence peak, in the $s_{\pm}$ state, where no
coherence effects take place, strong coupling is not really important. In
Fig.~\ref{fig3}(a) we show the effect of impurities in the Born limit. We have
found that for an impurity scattering of the order of $0.4 T_{c0}$, where
$T_{c0}$ is the transition temperature in the absence of impurities, there is a
moderate suppression of $T_{c}$ (less than 20\%). More importantly, the strong
deviation from exponential behavior in $1/T_{1}T$ appeared. Above $\sim 0.2
T_{c}$ the dependence can be well represented by a power law, but with an
exponent closer to $5.5$ for clean, and $4$ for dirty samples (as opposed to
the experimentally observed $1.5$ to $2$). Further increase of the scattering
rate leads to a too strong suppression of transition temperature and a too
large relaxation rate right below $T_{c}$. Thus, impurity scattering in the
Born limit can not fully explain the NMR data.

As shown in Fig.~\ref{fig3}(b), the unitary limit also does not reproduce the
experimental data (represented by the approximate behavior $1/T_{1}T \propto
T^{2}$) neither for small nor for large impurity concentrations, and even
predicts a slight non-monotonicity for small $\gamma=0.07 \Delta_{0}$. Here,
$\Delta_{0}$ is the low temperature value of the energy gap without impurity
scattering.
\begin{figure}[tbp]
\includegraphics[width=0.99\linewidth]{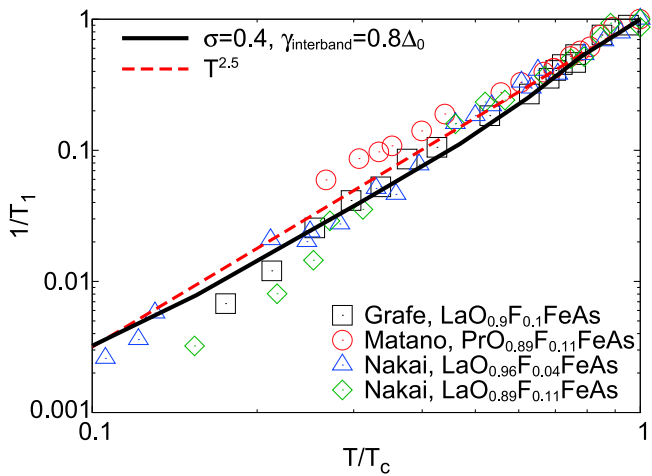}
\caption{(color online) Calculated spin-lattice relaxation rate for the
$s_{\pm}$ superconducting state together with experimental $1/T_{1}$
data from several groups, as indicated. $T^{2.5}$ trend is also shown.}
\label{figIntermediate}
\end{figure}

On the other hand, the experimental results can be reproduced very well if one
assumes the intermediate regime of impurity scattering.
Fig.~\ref{figIntermediate} shows various experimental data
\cite{NMR1,NMR2,NMR3} together with our calculations for the $s_{\pm}$ gap with
$\sigma$ taken as 0.4 and interband $\gamma$ taken as $0.8 \Delta_{0}$. This
$\gamma$ corresponds to a relatively dirty superconductivity, but the effect of
interband scattering on $T_{c}$, for given $\gamma$, in this $s_{\pm}$ state is
smaller than would be effected by intraband scattering of the same magnitude
\cite{kulic,senga}. Besides, as mentioned above, a distribution of $\sigma$'s
will lead to a similar broadening of the DOS for smaller $\gamma$.  We observe
again that the $s_{\pm}$ state exhibits no coherence peak. As opposed to the
Born and unitary limits, intermediate-$\sigma$ scattering is capable of
reproducing the experimental behavior, usually described as cubic, but in fact
probably closer to $T^{2.5}$ in $1/T_{1}$. Note that there is no universality
after the $2.5$ power of $T$, it is simply result of a fitting.

We want to emphasize that this analysis does not \textit{prove} that the origin
of the power law behavior is dirty-limit intermediate-$\sigma$ scattering in an
$s_{\pm}$ state. It is fairly possible that more complex physics, possibly
related to coexistence of superconductivity and spin density wave order, plays
a role. But it clearly demonstrates that such a behavior does not prove
existence of gap nodes on the Fermi surface.

To summarize, we have shown that the lack of a coherence peak is very naturally
explained in the framework of the $s_{\pm}$ superconducting state even in the
clean limit, and even more so in the presence of impurities. However, a
\textit{clean} $s_{\pm}$ superconductor would show an exponential decay of the
relaxation rate $1/T_{1}$ below $T_{c}$, contrary to what has been observed in
NMR experiments. Strong coupling effects and impurity scattering in Born
approximation transform this exponential behavior into a power law-like for
temperatures $T \gtrsim 0.2 T_{c}$, but it is difficult to reproduce the actual
experimental temperature dependence. On the other hand, an intermediate-limit
scattering (neither Born nor unitary), can reproduce experimental observations
rather closely. While we did not address in this paper any effects that strong
scattering may have on the other physical properties (this is left for future
publications), we want to emphasize that there is an important difference
between the scattering effect on the properties related to $\mathbf{q}=0$
response (penetration depth, tunneling, specific heat) and $1/T_{1}$ that
probes mainly the $\mathbf{q} \sim \mathbf{Q}$ response.

We would like to thank A. Chubukov, I. Eremin, K. Ishida, and G.-q. Zheng
for useful discussions.

\textit{Note added} As this paper was being finalized, we became aware of a
work by Chubukov \textit{et al.} \cite{Chubukov} who arrived at similar
conclusions using a different approach. After submission of our paper, Bang and
Choi \cite{Bang} reported a similar, but independent research, again reaching
the same conclusion as ours.


\begin{thebibliography}{99}
\bibitem{gaps} In the weak coupling limit, the gap ratio at $T \rightarrow
    T_{c}$ in the $s_{\pm}$ model is $\sqrt{N_{1}/N_{2}}$, where $N_{i}$ are
    densities of states. At $T \rightarrow 0$, $\Delta_{2}/\Delta_{1} \approx
    \sqrt{N_{1}/N_{2}} [1+\lambda \log (N_{1}/N_{2})/4]$, where $\lambda$ is
    the effective coupling constant (neglecting the intraband coupling). LDA
    calculations yield $N_{e}/N_{h} < 1.2$, therefore $\Delta_{1}$ and
    $\Delta_{2}$ differ by less than 10\%.
\end{thebibliography}
\end{document}